\documentclass[12pt]{article}

\usepackage{jheppub, bm, wrapfig,float,array}
\usepackage[utf8]{inputenc}
\numberwithin{equation}{section}
\renewcommand\afterTocRuleSpace{\bigskip}

\usepackage{amsfonts}
\usepackage{amsmath}
\usepackage{color}
\usepackage{graphicx}
\usepackage{float}
\usepackage[T1]{fontenc}
\usepackage[utf8]{inputenc}
\usepackage{alphabeta}
\usepackage{fancyvrb}
\usepackage[dvipsnames]{xcolor}
\usepackage{bold-extra}
\usepackage{lmodern}

\usepackage{tikz}
\usetikzlibrary{arrows}
\usetikzlibrary{shapes.geometric,calc,arrows, positioning,shapes.misc,decorations.markings}
\tikzset{
  big arrow/.style={
    decoration={markings,mark=at position 1 with {\arrow[scale=2,#1]{>}}},
    postaction={decorate},
    shorten >=0.4pt},
  big arrow/.default=black}

\newcommand{\bea}{\begin{eqnarray}}
\newcommand{\eea}{\end{eqnarray}}
\newcommand{\be}{\begin{equation}}
\newcommand{\ee}{\end{equation}}
\newcommand{\bit}{\begin{itemize}}
\newcommand{\eit}{\end{itemize}}
\newcommand{\ben}{\begin{enumerate}}
\newcommand{\een}{\end{enumerate}}

\renewcommand{\ni}{\noindent}

\newcommand{\half}{\frac{1}{2}}

\newcommand{\cC}{\mathcal{C}}

\newcommand{\cM}{\mathcal{M}}
\newcommand{\cN}{\mathcal{N}}

\newcommand{\cS}{\mathcal{S}}

\newcommand{\F}{\mathsf{F}}
\renewcommand{\S}{\mathsf{S}}

\newcommand{\fe}{\mathfrak{e}}
\newcommand{\ff}{\mathfrak{f}}
\newcommand{\fg}{\mathfrak{g}}

\newcommand{\su}{\mathfrak{su}}
\renewcommand{\sp}{\mathfrak{sp}}
\newcommand{\so}{\mathfrak{so}}

\newcommand{\ubf}[1]{\underline{\bf #1}}

\title{Do all $5d$ SCFTs descend from $6d$ SCFTs?}

\author{Lakshya Bhardwaj}

\affiliation{Department of Physics, Harvard University\\17 Oxford St, Cambridge, MA 02138, USA}

\abstract{We present examples of $5d$ SCFTs that serve as counter-examples to a recently actively studied conjecture according to which it should be possible to obtain all $5d$ SCFTs by integrating out BPS particles from $6d$ SCFTs compactified on a circle. We further observe that it is possible to obtain these $5d$ SCFTs from $6d$ SCFTs if one allows integrating out BPS strings as well. Based on this observation, we propose a revised version of the conjecture according to which it should be possible to obtain all $5d$ SCFTs by integrating out both BPS particles and BPS strings from $6d$ SCFTs compactified on a circle. We describe a general procedure to integrate out BPS strings from a $5d$ theory once a geometric description of the $5d$ theory is given. We also discuss the consequences of the revised conjecture for the classification program of $5d$ SCFTs.
}

\begin{document}

\maketitle

\section{Introduction and Discussion} \label{intro}
Recent work on the classification of $5d$ SCFTs was initiated by \cite{Jefferson:2017ahm,Jefferson:2018irk}. In particular, \cite{Jefferson:2018irk} provided significant evidence in support of a conjecture which formed the foundation for the subsequent work on the classification of $5d$ SCFTs \cite{Bhardwaj:2018yhy,Bhardwaj:2018vuu,Apruzzi:2018nre,Apruzzi:2019vpe,Apruzzi:2019opn,Bhardwaj:2019fzv,Bhardwaj:2019jtr}\footnote{See \cite{DelZotto:2017pti,Hayashi:2018bkd,Hayashi:2018lyv,Closset:2018bjz,Hayashi:2019yxj,Fluder:2019szh,Kim:2019dqn,Uhlemann:2019ypp,Bhardwaj:2019ngx,Apruzzi:2019enx,Saxena:2019wuy} for other related recent work on $5d$ SCFTs and $5d$ gauge theories.}. According to this conjecture, it should be possible to obtain all $5d$ SCFTs by performing RG flows on $6d$ SCFTs compactified on a circle. More specifically, it was assumed that these RG flows take the form of integrating out BPS particles from the $5d$ theories obtained by compactifying $6d$ SCFTs on a circle (often called $5d$ KK theories).

To understand the motivation behind the proposal of this conjecture, consider the example of a $5d$ $\cN=1$ pure gauge theory with gauge algebra $\fg$. Taking the strong coupling limit of this theory leads us to the conformal point of a $5d$ SCFT. This theory can be obtained by integrating out the adjoint hypermultiplet from $5d$ $\cN=1$ gauge theory with gauge algebra $\fg$ and a hypermultiplet transforming in adjoint representation. The latter $5d$ gauge theory can be obtained by compactifying a $6d$ $\cN=(2,0)$ SCFT on a circle of finite radius $R$ with the radius controlling the gauge coupling of the $5d$ gauge theory. The BPS particles being integrated out under this RG flow are the ones produced by the hyper in the adjoint representation once we move onto the Coulomb branch of the gauge theory.

In this work, we will present several examples of $5d$ SCFTs which cannot be obtained from $5d$ KK theories by integrating out BPS particles alone. The mass deformations of these $5d$ SCFTs admit certain Coulomb branch phases that can be described in terms of the following $5d$ $\cN=1$ gauge theories (deformed by their mass parameters and Coulomb branch moduli):
\bit
\item $\ff_4$ with $1\le n\le 3$ hypers in ``fundamental'' representation, that is, the irreducible representation of dimension $\mathbf{26}$.
\item $\fe_6$ with $1\le n\le 4$ hypers in ``fundamental'' representation, that is, the irreducible representation of dimension $\mathbf{27}$.
\item $\fe_7$ with $1\le n\le 6$ half-hypers in ``fundamental'' representation, that is, the irreducible representation of dimension $\mathbf{56}$.
\eit
These gauge theories can be constructed by compactifying M-theory on a Calabi-Yau threefold. We will show in Section \ref{T} that there is a point in the Kahler moduli space of the Calabi-Yau threefold where all of the compact 2-cycles and 4-cycles of the Calabi-Yau threefold shrink simultaneously to zero volume, thus giving rise to a $5d$ SCFT decoupled from gravity and other stringy physics. This argument establishes the existence of the above mentioned $5d$ SCFTs.

We claim that these $5d$ SCFTs cannot be obtained by integrating out BPS particles from a $5d$ KK theory. The argument proceeds in two steps:
\ben 
\item Let us first argue that for a $5d$ SCFT admitting a description as a $5d$ gauge theory, integrating in a BPS particle always corresponds to integrating in a matter hypermultiplet into the corresponding gauge theory.\\
For this argument, we use the M-theory construction of the $5d$ SCFT\footnote{Here we use the M-theory construction only to phrase the argument in a familiar language. One can phrase the argument in a purely field-theoretic language too.}. A BPS particle being integrated into this $5d$ SCFT arises as a compact 2-cycle $C$ living in a \emph{non-compact 4-cycle} $N$ and intersecting all the compact 4-cycles $S_i$ transversely. The flop of $C$ corresponds to the process of integrating in the BPS particle into the $5d$ SCFT, whereas decompactifying $C$ inside $N$ completely integrates it out. On the other hand, the Coulomb branch phase transitions of $5d$ SCFT are implemented by flopping compact 2-cycles living inside \emph{compact 4-cycles} $S_i$ in the Calabi-Yau threefold. Thus, the flops corresponding to addition of a BPS particle and the flops corresponding to Coulomb branch phase transitions commute. Consequently, the addition of a BPS particle into any phase of the $5d$ SCFT can always be thought as the addition of a BPS particle into a gauge-theoretic phase of the $5d$ SCFT described by the corresponding $5d$ gauge theory.\\
From the point of view of the gauge-theoretic phase, the compact 2-cycle $C$ can be combined with the compact 2-cycles $f_i$ (living inside $S_i$) corresponding to W-bosons of the gauge algebra $\fg$. Together, these form the BPS particles corresponding to a hypermultiplet charged in a representation $R$ of $\fg$. The Dynkin coefficients of the highest weight of $R$ are identified with the intersection numbers $C\cdot S_i$.
\item We can now use the results of \cite{Jefferson:2017ahm} according to which a $5d$ gauge theory obtained after adding a hypermultiplet to any of the following $5d$ gauge theories
\begin{align}
\ff_4&+3\F\label{T1}\\
\fe_6&+4\F\label{T2}\\
\fe_7&+3\F\label{T3}\\
\fe_7&+\frac52\F\label{T4}
\end{align}
(where $\F$ denotes a full hyper in fundamental representation) cannot describe a $5d$ SCFT or a $5d$ KK theory. Similarly, a $5d$ gauge theory obtained after adding a hypermultiplet in a representation other than the fundamental representation to any of the following $5d$ gauge theories
\begin{align}
\ff_4&+n\F~\text{ for }n<3\label{T5}\\
\fe_6&+n\F~\text{ for }n<4\label{T6}\\
\fe_7&+\frac n2\F~\text{ for }n<5\label{T7}
\end{align}
cannot describe a $5d$ SCFT or a $5d$ KK theory.
\een
We note that the above argument applies irrespective of whether the classification of $6d$ SCFTs \cite{Heckman:2015bfa,Bhardwaj:2019hhd}\footnote{See \cite{Heckman:2013pva,Bhardwaj:2015xxa,Bhardwaj:2018jgp,Seiberg:1996qx,Danielsson:1997kt} for other work related to the classification of $6d$ SCFTs.} is complete or not.

We will show in Section \ref{T} that the theories (\ref{T1}---\ref{T7}) can be obtained from $5d$ KK theories if one allows RG flows integrating out BPS strings as well. Such RG flows correspond to decompactifying compact 4-cycles in the Calabi-Yau threefold used in the corresponding M-theory construction. The BPS strings being integrated out during this process are the ones produced by compactifying M5 branes on the 4-cycles being decompactified. We will discuss the general geometric conditions under which such a decompactification process can consistently take place in Section \ref{decouple}.

The theories (\ref{T1}---\ref{T7}) can be obtained via such a decompactification process applied to $5d$ KK theories obtained by untwisted compactification of following $6d$ SCFTs\footnote{Here we are denoting the $6d$ SCFTs by the data of their F-theory construction.}:
\begin{align}
&\text{$\ff_4$ on $-k$ curve with $2\le k\le 5$}\label{6T1}\\
&\text{$\fe_6$ on $-k$ curve with $2\le k\le 6$}\label{6T2}\\
&\text{$\fe_7$ on $-k$ curve with $2\le k\le 8$}\label{6T3}
\end{align}

We are thus led to the conjecture that:
\begin{center}
\fbox{
\begin{minipage}[c][2cm][c]{15cm}
\ubf{Conjecture 1}: All $5d$ SCFTs can be obtained (on their mass-deformed Coulomb branch) by consistently integrating out BPS particles and BPS strings from the mass-deformed Coulomb branches of $5d$ KK theories.
\end{minipage}
}
\end{center}

\vspace{2mm}

The above revised conjecture has a negative consequence for the program of classifying $5d$ SCFTs based on the analysis of RG flows of $5d$ KK theories. To understand this, let us notice that the RG flows corresponding to removal of BPS particles are \emph{rank-preserving}, that is they do not change the rank of the $5d$ theory, which can be identified with the number of fundamental BPS strings\footnote{We define fundamental BPS strings to be those BPS strings that do not arise as bound states of other BPS strings.} found on the mass-deformed Coulomb branch of the $5d$ theory. On the other hand, the RG flows corresponding to removal of BPS strings are \emph{rank-lowering}. If all $5d$ SCFTs could be obtained by only integrating out BPS particles from $5d$ KK theories, one would only need to analyze RG flows of $5d$ KK theories of rank $n$ in order to fully classify $5d$ SCFTs of rank $n$. This was the logic behind the classification of $5d$ SCFTs upto rank three pursued in \cite{Jefferson:2018irk,Bhardwaj:2019jtr}. However, since we also need to integrate out BPS strings, we need to actually analyze RG flows of $5d$ KK theories of ranks greater than or equal to $n$ in order to fully classify $5d$ SCFTs of rank $n$. For example, we might encounter the situation in which a $5d$ SCFT of rank $n$ only arises by removal of a BPS string from a $5d$ SCFT of rank $n+1$, which only arises by removal of a BPS string from a $5d$ SCFT of rank $n+2$, and so on until we reach a $5d$ SCFT of rank $n+p$ which only arises by removal of a BPS string from a $5d$ KK theory of rank $n+p+1$. Thus, \emph{rank} is no longer a good notion to organize the classification of $5d$ SCFTs. To remedy the situation, we make the following conjecture:
\begin{center}
\fbox{
\begin{minipage}[c][2cm][c]{15cm}
\ubf{Conjecture 2}: A rank $n$ $5d$ SCFT can be obtained either by successively integrating out BPS particles from a rank $n$ $5d$ KK theory, or by successively integrating out BPS particles after integrating out a BPS string from a rank $n+1$ $5d$ KK theory.
\end{minipage}
}
\end{center}
Evidence for this conjecture will be provided in \cite{Bhardwaj:2020gyu} where it will be shown that every $5d$ gauge theory (with a simple gauge algebra) that is expected to arise on the mass-deformed Coulomb branch of a $5d$ SCFT (based on the analysis of \cite{Jefferson:2017ahm} and a few other conditions) can be obtained from a $5d$ KK theory by using only the RG flows mentioned in {\bf Conjecture 2}. 

It would be an interesting future direction to explore if {\bf Conjecture 2} leads to an extension of the classification of $5d$ SCFTs presented in \cite{Bhardwaj:2019jtr}. This would require studying all the possible decompactifications of a single 4-cycle in the Calabi-Yau threefolds associated to $5d$ KK theories upto rank four. The general criteria for decompactifying a compact 4-cycle in a Calabi-Yau threefold is discussed in Section \ref{decouple}.

\section{Construction of the counter-examples}\label{T}
In this section, we will show that the $5d$ SCFTs described by the $5d$ gauge theories (\ref{T1}---\ref{T7}) can be obtained by integrating out a BPS string from the $5d$ KK theories obtained by untwisted compactification of $6d$ SCFTs (\ref{6T1}---\ref{6T3}). This construction will also allow us to exhibit the existence of a ray in the space of normalizable Kahler parameters for the Calabi-Yau threefolds associated to the theories (\ref{T1}---\ref{T7}), such that all of the compact 2-cycles and 4-cycles have non-negative volumes along the ray, and moreover at least one of the 4-cycles has strictly positive volume. The existence of such a ray implies that these Calabi-Yau threefolds describe $5d$ SCFTs \cite{Jefferson:2018irk}, with the ray becoming a part of the Coulomb branch of the $5d$ SCFTs, and the origin of the ray becoming the conformal point.

We will use the notation detailed in Section 5.2.1 of \cite{Bhardwaj:2019fzv} to describe Calabi-Yau threefolds throughout this paper.

Let us start by recalling the Calabi-Yau threefold associated to the untwisted compactification of the $6d$ SCFT carrying $\ff_4$ on $-k$ curve \cite{Bhardwaj:2018yhy,Bhardwaj:2018vuu,Bhardwaj:2019fzv}:
\be\label{f4k}
\begin{tikzpicture} [scale=1.9]
\node (v1) at (-4,0) {$\mathbf{4_{4-k}}$};
\node (v2) at (-2.9,0) {$\mathbf{3_{6-k}}$};
\node (v3) at (-0.7,0) {$\mathbf{2_{6}^{(5-k)+(5-k)}}$};
\draw  (v1) edge (v2);
\draw  (v2) edge (v3);
\node at (-4.1,0.6) {\scriptsize{$e$}};
\node at (-4.1,0.3) {\scriptsize{$e$}};
\node at (-3.6,0.1) {\scriptsize{$h$}};
\node (v6) at (3.3,0) {$\mathbf{1_{8}^{(5-k)+(5-k)}}$};
\node at (-3.3,0.1) {\scriptsize{$e$}};
\node at (-2.5,0.1) {\scriptsize{ $2h$}};
\node (v9) at (-4,0.9) {$\mathbf{0_{k-2}}$};
\draw  (v9) edge (v1);
\node at (-1.8,0.1) {\scriptsize{$e$ -$\sum x_i$-$\sum y_i$}};
\node at (2.3,0.1) {\scriptsize{$e$-$\sum y_i,f$-$x_i$}};
\node (u1) at (-0.7,-0.6) {\scriptsize{$5-k$}};
\draw (v3) .. controls (-0.9,-0.2) and (-1.4,-0.5) .. (u1);
\draw (v3) .. controls (-0.5,-0.2) and (0,-0.5) .. (u1);
\node at (-1.2,-0.3) {\scriptsize{$x_i$}};
\node at (-0.2,-0.3) {\scriptsize{$y_i$}};
\node at (0.6,0.1) {\scriptsize{$h+\sum(f$-$y_i),f$-$x_i$}};
\node (v4) at (1.5,0) {\scriptsize{$6-k$}};
\draw  (v3) edge (v4);
\draw  (v4) edge (v6);
\node (u2) at (3.3,-0.6) {\scriptsize{$5-k$}};
\draw (v6) .. controls (3.1,-0.2) and (2.6,-0.5) .. (u2);
\draw (v6) .. controls (3.5,-0.2) and (4,-0.5) .. (u2);
\node at (2.8,-0.3) {\scriptsize{$x_i$}};
\node at (3.8,-0.3) {\scriptsize{$y_i$}};
\end{tikzpicture}
\ee
for $2\le k\le 5$, and
\be\label{f41}
\begin{tikzpicture} [scale=1.9]
\node (v1) at (-3.7,0) {$\mathbf{4_{3}}$};
\node (v2) at (-2.7,0) {$\mathbf{3_{5}}$};
\node (v3) at (-0.7,0) {$\mathbf{2_{6}^{4+4}}$};
\draw  (v1) edge (v2);
\draw  (v2) edge (v3);
\node at (-4.5,0.1) {\scriptsize{$h$}};
\node at (-4,0.1) {\scriptsize{$e$}};
\node at (-3.4,0.1) {\scriptsize{$h$}};
\node (v6) at (2.8,0) {$\mathbf{1_{8}^{4+4}}$};
\node at (-3,0.1) {\scriptsize{$e$}};
\node at (-2.4,0.1) {\scriptsize{ $2h$}};
\node (v9) at (-4.8,0) {$\mathbf{0_{1}}$};
\draw  (v9) edge (v1);
\node at (-1.5,0.1) {\scriptsize{$e$ -$\sum x_i$-$\sum y_i$}};
\node at (2,0.1) {\scriptsize{$e$-$\sum y_i,f$-$x_i$}};
\node (u1) at (-0.7,-0.6) {\scriptsize{$4$}};
\draw (v3) .. controls (-0.9,-0.2) and (-1.4,-0.5) .. (u1);
\draw (v3) .. controls (-0.5,-0.2) and (0,-0.5) .. (u1);
\node at (-1.2,-0.3) {\scriptsize{$x_i$}};
\node at (-0.2,-0.3) {\scriptsize{$y_i$}};
\node at (0.3,0.1) {\scriptsize{$h+\sum(f$-$y_i),f$-$x_i$}};
\node (v4) at (1.2,0) {\scriptsize{$5$}};
\draw  (v3) edge (v4);
\draw  (v4) edge (v6);
\node (u2) at (2.8,-0.6) {\scriptsize{$4$}};
\draw (v6) .. controls (2.6,-0.2) and (2.1,-0.5) .. (u2);
\draw (v6) .. controls (3,-0.2) and (3.5,-0.5) .. (u2);
\node at (2.3,-0.3) {\scriptsize{$x_i$}};
\node at (3.3,-0.3) {\scriptsize{$y_i$}};
\end{tikzpicture}
\ee
for $k=1$. These Calabi-Yau threefolds admit a special ray in the space of normalizable Kahler parameters where the Kahler form $J$ can be written as
\be\label{rf4}
J=\phi\left(S_0+2S_4+3S_3+2S_2+S_1\right)
\ee
with $\phi\ge0$. The coefficients of $S_i$ in this ray are actually dual Coxeter labels associated to roots of the affine algebra $\ff_4^{(1)}$. Along this ray, all of the compact 2-cycles have non-negative volume and all the compact 4-cycles have zero volume, as the reader can explicitly check.

Now, let us send the volume of the curve $f$ in $S_0$ of (\ref{f4k}) to infinity, while keeping the volume of $e$ curve in $S_0$ finite. This decompactifies $S_0$ and we obtain the Calabi-Yau threefold
\be\label{F4k}
\begin{tikzpicture} [scale=1.9]
\node (v1) at (-4,0) {$\mathbf{4_{4-k}}$};
\node (v2) at (-2.9,0) {$\mathbf{3_{6-k}}$};
\node (v3) at (-0.7,0) {$\mathbf{2_{6}^{(5-k)+(5-k)}}$};
\draw  (v1) edge (v2);
\draw  (v2) edge (v3);
\node at (-3.6,0.1) {\scriptsize{$h$}};
\node (v6) at (3.3,0) {$\mathbf{1_{8}^{(5-k)+(5-k)}}$};
\node at (-3.3,0.1) {\scriptsize{$e$}};
\node at (-2.5,0.1) {\scriptsize{ $2h$}};
\node at (-1.8,0.1) {\scriptsize{$e$ -$\sum x_i$-$\sum y_i$}};
\node at (2.3,0.1) {\scriptsize{$e$-$\sum y_i,f$-$x_i$}};
\node (u1) at (-0.7,-0.6) {\scriptsize{$5-k$}};
\draw (v3) .. controls (-0.9,-0.2) and (-1.4,-0.5) .. (u1);
\draw (v3) .. controls (-0.5,-0.2) and (0,-0.5) .. (u1);
\node at (-1.2,-0.3) {\scriptsize{$x_i$}};
\node at (-0.2,-0.3) {\scriptsize{$y_i$}};
\node at (0.6,0.1) {\scriptsize{$h+\sum(f$-$y_i),f$-$x_i$}};
\node (v4) at (1.5,0) {\scriptsize{$6-k$}};
\draw  (v3) edge (v4);
\draw  (v4) edge (v6);
\node (u2) at (3.3,-0.6) {\scriptsize{$5-k$}};
\draw (v6) .. controls (3.1,-0.2) and (2.6,-0.5) .. (u2);
\draw (v6) .. controls (3.5,-0.2) and (4,-0.5) .. (u2);
\node at (2.8,-0.3) {\scriptsize{$x_i$}};
\node at (3.8,-0.3) {\scriptsize{$y_i$}};
\end{tikzpicture}
\ee
which describes the $5d$ gauge theory $\ff_4+(5-k)\F$ for $2\le k\le 5$. 

Since (\ref{F4k}) is a limit of (\ref{f4k}), and (\ref{f4k}) describes a theory which is UV complete without coupling to dynamical gravity, (\ref{F4k}) should also describe a theory which is UV complete without coupling to dynamical gravity. In fact, we claim that (\ref{F4k}) describes a $5d$ SCFT. To show this, we study the Calabi-Yau threefold (\ref{F4k}) along the ray 
\be\label{Rf4}
J=\phi\left(2S_4+3S_3+2S_2+S_1\right)
\ee
which is the ray (\ref{rf4}) with $S_0$ deleted. First of all, the volume of any 2-cycle or 4-cycle not intersecting $S_0$ will remain unchanged. Thus only the volumes of 2-cycles $e,f$ in $S_4$ and the volume of the 4-cycle $S_4$ will change. It can be checked that both the 2-cycles attain non-negative volume and $S_4$ attains strictly positive volume along (\ref{Rf4}). As a consequence, we have shown that $\ff_4+(5-k)\F$ describes a $5d$ SCFT for $2\le k\le 5$ as claimed in Section \ref{intro}.

On the other hand, an analogous RG flow is not possible for the $k=1$ case shown in (\ref{f41}). Decompactifying $f$ of $S_0$ while keeping $e$ of $S_0$ compact in (\ref{f41}) necessarily decompactifies $h$ of $S_0$ which is glued to $e$ of $S_4$, thus decompactifying $S_4$ in the process as well\footnote{In fact, following this reasoning, we can see that this process will decompactify all the other surfaces as well.}. Thus, we are unable to obtain an $\ff_4$ gauge theory from (\ref{f41}). This was as expected, since according to the analysis of \cite{Jefferson:2017ahm}, $\ff_4+4\F$ does not describe a $5d$ SCFT or a $5d$ KK theory. Thus $\ff_4+4\F$ either cannot be UV completed or requires coupling to dynamical gravity for consistent UV completion. We would find a contradiction if we were able to decompactify $S_0$ while keeping the rest of the Calabi-Yau threefold compact.

A similar argument works for $\fe_6$ and $\fe_7$ theories. For $\fe_6$ theories, we start with the Calabi-Yau threefold for untwisted compactification of the $6d$ SCFT carrying $\fe_6$ on $-k$ curve
\be
\begin{tikzpicture} [scale=1.9]
\node (v1) at (-4,0) {$\mathbf{0_{k-2}}$};
\node (v2) at (-2.8,0) {$\mathbf{6_{k-4}}$};
\node (v3) at (-1.6,0) {$\mathbf{3_{6-k}}$};
\node (v4) at (-1.6,2.4) {$\mathbf{5_{4}}$};
\draw  (v1) edge (v2);
\draw  (v2) edge (v3);
\node at (-3.6,-0.1) {\scriptsize{$e$}};
\node at (-0.5,-0.1) {\scriptsize{$e$}};
\node at (-3.2,-0.1) {\scriptsize{$h$}};
\node at (-2.4,-0.1) {\scriptsize{$e$}};
\node at (1.2,0.5) {\scriptsize{$x_i$-$y_i$}};
\node[rotate=0] at (0.4,-0.1) {\scriptsize{$h$}};
\node (v5) at (-0.8,0.6) {\scriptsize{$6-k$}};
\node at (-1.2,-0.1) {\scriptsize{$h$}};
\node (v8) at (-1.6,1.4) {$\mathbf{4_{8-k}^{6-k}}$};
\node (v6) at (-0.1,0) {$\mathbf{2_{8-k}^{6-k}}$};
\node (v7) at (1.5,0) {$\mathbf{1_{10-k}^{(6-k)+(6-k)}}$};
\draw  (v3) edge (v6);
\draw  (v6) edge (v7);
\draw  (v3) edge (v8);
\draw  (v8) edge (v4);
\node (v9) at (-0.3,0.8) {\scriptsize{$6-k$}};
\node (v10) at (-0.1,1.2) {\scriptsize{$6-k$}};
\node at (-2,-0.1) {\scriptsize{$e$}};
\node at (-1.7,0.3) {\scriptsize{$h$}};
\node at (0.8,-0.1) {\scriptsize{$e$}};
\node at (-1.1,1.3) {\scriptsize{$x_i$}};
\node [rotate=-20] at (0.4,0.6) {\scriptsize{$f$-$x_i$}};
\node at (-1.2,2.3) {\scriptsize{$f$}};
\node at (-1.7,1.1) {\scriptsize{$e$}};
\draw  (v8) edge (v5);
\draw  (v5) edge (v6);
\draw  (v8) edge (v9);
\draw  (v9) edge (v7);
\draw  (v4) edge (v10);
\draw  (v10) edge (v7);
\node at (-1.9,1.7) {\scriptsize{$h$-$\sum x_i$}};
\node at (-1.8,2.1) {\scriptsize{$e$}};
\node [rotate=0] at (-1.3,0.9) {\scriptsize{$f$-$x_i$}};
\node [rotate=0] at (-0.7,0.3) {\scriptsize{$f$-$x_i$}};
\end{tikzpicture}
\ee
for $2\le k\le 6$. The dual Coxeter labels for $\fe_6^{(1)}$ provide us with the ray
\be
J=\phi\left(S_0+S_1+S_5+2S_2+2S_4+2S_6+3S_3\right)
\ee
For $k\ge2$, it is possible to decompactify $S_0$ while keeping other $S_i$ compact by decompactifying $f$ of $S_0$. We obtain
\be\label{E6}
\begin{tikzpicture} [scale=1.9]
\node (v2) at (-2.8,0) {$\mathbf{6_{k-4}}$};
\node (v3) at (-1.6,0) {$\mathbf{3_{6-k}}$};
\node (v4) at (-1.6,2.4) {$\mathbf{5_{4}}$};
\draw  (v2) edge (v3);
\node at (-0.5,-0.1) {\scriptsize{$e$}};
\node at (-2.4,-0.1) {\scriptsize{$e$}};
\node at (1.2,0.5) {\scriptsize{$x_i$-$y_i$}};
\node[rotate=0] at (0.4,-0.1) {\scriptsize{$h$}};
\node (v5) at (-0.8,0.6) {\scriptsize{$6-k$}};
\node at (-1.2,-0.1) {\scriptsize{$h$}};
\node (v8) at (-1.6,1.4) {$\mathbf{4_{8-k}^{6-k}}$};
\node (v6) at (-0.1,0) {$\mathbf{2_{8-k}^{6-k}}$};
\node (v7) at (1.5,0) {$\mathbf{1_{10-k}^{(6-k)+(6-k)}}$};
\draw  (v3) edge (v6);
\draw  (v6) edge (v7);
\draw  (v3) edge (v8);
\draw  (v8) edge (v4);
\node (v9) at (-0.3,0.8) {\scriptsize{$6-k$}};
\node (v10) at (-0.1,1.2) {\scriptsize{$6-k$}};
\node at (-2,-0.1) {\scriptsize{$e$}};
\node at (-1.7,0.3) {\scriptsize{$h$}};
\node at (0.8,-0.1) {\scriptsize{$e$}};
\node at (-1.1,1.3) {\scriptsize{$x_i$}};
\node [rotate=-20] at (0.4,0.6) {\scriptsize{$f$-$x_i$}};
\node at (-1.2,2.3) {\scriptsize{$f$}};
\node at (-1.7,1.1) {\scriptsize{$e$}};
\draw  (v8) edge (v5);
\draw  (v5) edge (v6);
\draw  (v8) edge (v9);
\draw  (v9) edge (v7);
\draw  (v4) edge (v10);
\draw  (v10) edge (v7);
\node at (-1.9,1.7) {\scriptsize{$h$-$\sum x_i$}};
\node at (-1.8,2.1) {\scriptsize{$e$}};
\node [rotate=0] at (-1.3,0.9) {\scriptsize{$f$-$x_i$}};
\node [rotate=0] at (-0.7,0.3) {\scriptsize{$f$-$x_i$}};
\end{tikzpicture}
\ee
which describes the $5d$ gauge theory $\fe_6+(6-k)\F$ for $2\le k\le 6$. The ray
\be
J=\phi\left(S_1+S_5+2S_2+2S_4+2S_6+3S_3\right)
\ee
which implies that the Calabi-Yau threefold (\ref{E6}) describes a $5d$ SCFT for $2\le k\le 6$, thus establishing that (\ref{T2}) and (\ref{T6}) are indeed $5d$ SCFTs.

For $\fe_7+n\F$ theories, we start with the Calabi-Yau threefold for the untwisted compactification of the $6d$ SCFT carrying $\fe_7$ on $-k$ curve
\be
\begin{tikzpicture} [scale=1.9]
\node (v1) at (-4,0) {$\mathbf{6_{2m-4}}$};
\node (v2) at (-2.8,0) {$\mathbf{5_{2m-6}}$};
\node (v3) at (-1.3,0) {$\mathbf{4_{4-m}^{4-m}}$};
\node (v4) at (3.45,0) {$\mathbf{1_{18-3m}^{4-m}}$};
\draw  (v1) edge (v2);
\draw  (v2) edge (v3);
\node at (-4.75,-0.1) {\scriptsize{$e$ }};
\node at (-4.4,-0.1) {\scriptsize{$h$ }};
\node at (-3.55,-0.1) {\scriptsize{$e$ }};
\node (v6) at (0.2,0) {$\mathbf{3_{6-m}^{4-m}}$};
\node (v7) at (1.7,0) {$\mathbf{2_{8-m}^{4-m}}$};
\draw  (v3) edge (v6);
\draw  (v6) edge (v7);
\node at (-3.2,-0.1) {\scriptsize{$h$}};
\node at (-0.15,-0.1) {\scriptsize{$e$}};
\node at (0.6,-0.1) {\scriptsize{$h$}};
\node at (-2.3,-0.1) {\scriptsize{ $e$}};
\node (v9) at (-5.2,0) {$\mathbf{0_{2m-2}}$};
\draw  (v9) edge (v1);
\draw  (v7) edge (v4);
\node at (-1.8,-0.1) {\scriptsize{$e$-$\sum x_i$}};
\node at (-1.4,0.3) {\scriptsize{$h$}};
\node at (1.35,-0.1) {\scriptsize{$e$}};
\node at (-0.9,-0.1) {\scriptsize{$h$}};
\node (v10) at (-1.3,2.2) {$\mathbf{7_{6-m}^{4-m}}$};
\node at (2.35,-0.1) {\scriptsize{$h$+($4$-$m$)$f$}};
\node at (2.95,-0.1) {\scriptsize{$e$}};
\node at (-1.4,1.8) {\scriptsize{$e$}};
\node at (1.4,0.4) {\scriptsize{$f$-$x_i$}};
\draw  (v10) edge (v3);
\node (u1) at (-0.4,0.9) {\scriptsize{$4-m$}};
\node (v5) at (0.4,0.9) {\scriptsize{$4-m$}};
\node (v8) at (1.4,1) {\scriptsize{$4-m$}};
\node at (-0.5,2) {\scriptsize{$f$-$x_i$}};
\node at (3,0.4) {\scriptsize{$f$-$x_i$}};
\node at (-0.5,1.7) {\scriptsize{$x_i$}};
\node at (-1,1.5) {\scriptsize{$f$-$x_i$}};
\node at (-0.3,0.3) {\scriptsize{$f$-$x_i$}};
\node at (-1.2,-0.4) { \scriptsize{ $ f$-$x_i$} };
\node at (3.5,-0.4) {\scriptsize{ $x_i$} };
\node at (2.9,-0.5) {\scriptsize{ $ f$-$x_i$} };
\node at (0.4,-0.4) {\scriptsize{ $ f$-$x_i$} };
\draw  (v10) edge (u1);
\draw  (u1) edge (v6);
\draw  (v10) edge (v5);
\draw  (v5) edge (v7);
\draw  (v10) edge (v8);
\draw  (v8) edge (v4);
\node (w1) at (1.8,-0.6) {\scriptsize{$4-m$}};
\node (w2) at (1.7,-1.1) {\scriptsize{$4-m$}};
\draw (v3) .. controls (-0.9,-0.6) and (0.6,-1) .. (w2);
\draw (w2) .. controls (2.5,-1.1) and (3.3,-0.8) .. (v4);
\draw (v6) .. controls (0.2,-0.2) and (1,-0.5) .. (w1);
\draw (w1) .. controls (2.7,-0.5) and (3.3,-0.2) .. (v4);
\end{tikzpicture}
\ee
for $k=2m$ and $1\le m\le 4$. Decompactifying $f$ of $S_0$ leads to the Calabi-Yau threefold
\be
\begin{tikzpicture} [scale=1.9]
\node (v1) at (-4,0) {$\mathbf{6_{2m-4}}$};
\node (v2) at (-2.8,0) {$\mathbf{5_{2m-6}}$};
\node (v3) at (-1.3,0) {$\mathbf{4_{4-m}^{4-m}}$};
\node (v4) at (3.45,0) {$\mathbf{1_{18-3m}^{4-m}}$};
\draw  (v1) edge (v2);
\draw  (v2) edge (v3);
\node at (-3.55,-0.1) {\scriptsize{$e$ }};
\node (v6) at (0.2,0) {$\mathbf{3_{6-m}^{4-m}}$};
\node (v7) at (1.7,0) {$\mathbf{2_{8-m}^{4-m}}$};
\draw  (v3) edge (v6);
\draw  (v6) edge (v7);
\node at (-3.2,-0.1) {\scriptsize{$h$}};
\node at (-0.15,-0.1) {\scriptsize{$e$}};
\node at (0.6,-0.1) {\scriptsize{$h$}};
\node at (-2.3,-0.1) {\scriptsize{ $e$}};
\draw  (v7) edge (v4);
\node at (-1.8,-0.1) {\scriptsize{$e$-$\sum x_i$}};
\node at (-1.4,0.3) {\scriptsize{$h$}};
\node at (1.35,-0.1) {\scriptsize{$e$}};
\node at (-0.9,-0.1) {\scriptsize{$h$}};
\node (v10) at (-1.3,2.2) {$\mathbf{7_{6-m}^{4-m}}$};
\node at (2.35,-0.1) {\scriptsize{$h$+($4$-$m$)$f$}};
\node at (2.95,-0.1) {\scriptsize{$e$}};
\node at (-1.4,1.8) {\scriptsize{$e$}};
\node at (1.4,0.4) {\scriptsize{$f$-$x_i$}};
\draw  (v10) edge (v3);
\node (u1) at (-0.4,0.9) {\scriptsize{$4-m$}};
\node (v5) at (0.4,0.9) {\scriptsize{$4-m$}};
\node (v8) at (1.4,1) {\scriptsize{$4-m$}};
\node at (-0.5,2) {\scriptsize{$f$-$x_i$}};
\node at (3,0.4) {\scriptsize{$f$-$x_i$}};
\node at (-0.5,1.7) {\scriptsize{$x_i$}};
\node at (-1,1.5) {\scriptsize{$f$-$x_i$}};
\node at (-0.3,0.3) {\scriptsize{$f$-$x_i$}};
\node at (-1.2,-0.4) { \scriptsize{ $ f$-$x_i$} };
\node at (3.5,-0.4) {\scriptsize{ $x_i$} };
\node at (2.9,-0.5) {\scriptsize{ $ f$-$x_i$} };
\node at (0.4,-0.4) {\scriptsize{ $ f$-$x_i$} };
\draw  (v10) edge (u1);
\draw  (u1) edge (v6);
\draw  (v10) edge (v5);
\draw  (v5) edge (v7);
\draw  (v10) edge (v8);
\draw  (v8) edge (v4);
\node (w1) at (1.8,-0.6) {\scriptsize{$4-m$}};
\node (w2) at (1.7,-1.1) {\scriptsize{$4-m$}};
\draw (v3) .. controls (-0.9,-0.6) and (0.6,-1) .. (w2);
\draw (w2) .. controls (2.5,-1.1) and (3.3,-0.8) .. (v4);
\draw (v6) .. controls (0.2,-0.2) and (1,-0.5) .. (w1);
\draw (w1) .. controls (2.7,-0.5) and (3.3,-0.2) .. (v4);
\end{tikzpicture}
\ee
which describes the $5d$ gauge theory $\fe_7+(4-m)\F$ for $1\le m\le 4$. The dual Coxeter labels for $\fe_7^{(1)}$ suggest studying the Calabi-Yau threefold along
\be\label{D}
J=\phi\left(S_1+2S_2+2S_6+2S_7+3S_3+3S_5+4S_4\right)
\ee
where we find that the Calabi-Yau threefold describes a $5d$ SCFT for all $1\le m\le 4$.

For $\fe_7+\left(n+\half\right)\F$ theories, we start with the Calabi-Yau threefold for the untwisted compactification of $6d$ SCFT carrying $\fe_7$ on $-k$ curve
\be
\begin{tikzpicture} [scale=1.9]
\node (v1) at (-4,0) {$\mathbf{6_{2m-5}}$};
\node (v2) at (-2.8,0) {$\mathbf{5_{2m-7}}$};
\node (v3) at (-1.3,0) {$\mathbf{4_{5-m}^{4-m}}$};
\node (v4) at (3.45,0) {$\mathbf{1_{20-3m}^{4-m}}$};
\draw  (v1) edge (v2);
\draw  (v2) edge (v3);
\node at (-4.75,-0.1) {\scriptsize{$e$ }};
\node at (-4.4,-0.1) {\scriptsize{$h$ }};
\node at (-3.55,-0.1) {\scriptsize{$e$ }};
\node (v6) at (0.2,0) {$\mathbf{3_{6-m}^{(4-m)+2}}$};
\node (v7) at (1.7,0) {$\mathbf{2_{8-m}^{4-m}}$};
\draw  (v3) edge (v6);
\draw  (v6) edge (v7);
\node at (-3.2,-0.1) {\scriptsize{$h$}};
\node at (-0.4,-0.1) {\scriptsize{$e$-$y_1$}};
\node at (0.8,-0.1) {\scriptsize{$h$}};
\node at (-2.3,-0.1) {\scriptsize{ $e$}};
\node (v9) at (-5.2,0) {$\mathbf{0_{2m-3}}$};
\draw  (v9) edge (v1);
\draw  (v7) edge (v4);
\node at (-1.8,-0.1) {\scriptsize{$e$-$\sum x_i$}};
\node at (-1.4,0.3) {\scriptsize{$h$}};
\node at (1.3,-0.1) {\scriptsize{$e$}};
\node at (-0.9,-0.1) {\scriptsize{$h$}};
\node (v10) at (-1.3,2.2) {$\mathbf{7_{7-m}^{4-m}}$};
\node at (2.35,-0.1) {\scriptsize{$h$+($5$-$m$)$f$}};
\node at (2.95,-0.1) {\scriptsize{$e$}};
\node at (-1.4,1.8) {\scriptsize{$e$}};
\node at (1.4,0.4) {\scriptsize{$f$-$x_i$}};
\draw  (v10) edge (v3);
\node (u1) at (-0.4,0.9) {\scriptsize{$5-m$}};
\node (v5) at (0.4,0.9) {\scriptsize{$4-m$}};
\node (v8) at (1.4,1) {\scriptsize{$4-m$}};
\node at (-0.5,2) {\scriptsize{$f$-$x_i$}};
\node at (3,0.4) {\scriptsize{$f$-$x_i$}};
\node at (-0.5,1.7) {\scriptsize{$x_i$}};
\node at (-1,1.4) {\scriptsize{$f$,$f$-$x_i$}};
\node at (-0.5,0.4) {\scriptsize{$y_1$-$y_2$,$f$-$x_i$}};
\node at (-1.2,-0.4) { \scriptsize{ $ f$-$x_i$} };
\node at (3.5,-0.4) {\scriptsize{ $x_i$} };
\node at (2.7,-0.6) {\scriptsize{$f$,$f$-$x_i$}};
\node at (0.2,-0.4) {\scriptsize{$f$-$y_1$-$y_2$,$f$-$x_i$}};
\draw  (v10) edge (u1);
\draw  (u1) edge (v6);
\draw  (v10) edge (v5);
\draw  (v5) edge (v7);
\draw  (v10) edge (v8);
\draw  (v8) edge (v4);
\node (w1) at (1.8,-0.6) {\scriptsize{$5-m$}};
\node (w2) at (1.7,-1.1) {\scriptsize{$4-m$}};
\draw (v3) .. controls (-0.9,-0.6) and (0.6,-1) .. (w2);
\draw (w2) .. controls (2.5,-1.1) and (3.3,-0.8) .. (v4);
\draw (v6) .. controls (0.2,-0.2) and (1,-0.5) .. (w1);
\draw (w1) .. controls (2.7,-0.5) and (3.3,-0.2) .. (v4);
\end{tikzpicture}
\ee
for $k=2m-1$ and $2\le m\le 4$. Decompactifying $f$ of $S_0$ leads to the Calabi-Yau threefold
\be
\begin{tikzpicture} [scale=1.9]
\node (v1) at (-4,0) {$\mathbf{6_{2m-5}}$};
\node (v2) at (-2.8,0) {$\mathbf{5_{2m-7}}$};
\node (v3) at (-1.3,0) {$\mathbf{4_{5-m}^{4-m}}$};
\node (v4) at (3.45,0) {$\mathbf{1_{20-3m}^{4-m}}$};
\draw  (v1) edge (v2);
\draw  (v2) edge (v3);
\node at (-3.55,-0.1) {\scriptsize{$e$ }};
\node (v6) at (0.2,0) {$\mathbf{3_{6-m}^{(4-m)+2}}$};
\node (v7) at (1.7,0) {$\mathbf{2_{8-m}^{4-m}}$};
\draw  (v3) edge (v6);
\draw  (v6) edge (v7);
\node at (-3.2,-0.1) {\scriptsize{$h$}};
\node at (-0.4,-0.1) {\scriptsize{$e$-$y_1$}};
\node at (0.8,-0.1) {\scriptsize{$h$}};
\node at (-2.3,-0.1) {\scriptsize{ $e$}};
\draw  (v7) edge (v4);
\node at (-1.8,-0.1) {\scriptsize{$e$-$\sum x_i$}};
\node at (-1.4,0.3) {\scriptsize{$h$}};
\node at (1.3,-0.1) {\scriptsize{$e$}};
\node at (-0.9,-0.1) {\scriptsize{$h$}};
\node (v10) at (-1.3,2.2) {$\mathbf{7_{7-m}^{4-m}}$};
\node at (2.35,-0.1) {\scriptsize{$h$+($5$-$m$)$f$}};
\node at (2.95,-0.1) {\scriptsize{$e$}};
\node at (-1.4,1.8) {\scriptsize{$e$}};
\node at (1.4,0.4) {\scriptsize{$f$-$x_i$}};
\draw  (v10) edge (v3);
\node (u1) at (-0.4,0.9) {\scriptsize{$5-m$}};
\node (v5) at (0.4,0.9) {\scriptsize{$4-m$}};
\node (v8) at (1.4,1) {\scriptsize{$4-m$}};
\node at (-0.5,2) {\scriptsize{$f$-$x_i$}};
\node at (3,0.4) {\scriptsize{$f$-$x_i$}};
\node at (-0.5,1.7) {\scriptsize{$x_i$}};
\node at (-1,1.4) {\scriptsize{$f$,$f$-$x_i$}};
\node at (-0.5,0.4) {\scriptsize{$y_1$-$y_2$,$f$-$x_i$}};
\node at (-1.2,-0.4) { \scriptsize{ $ f$-$x_i$} };
\node at (3.5,-0.4) {\scriptsize{ $x_i$} };
\node at (2.7,-0.6) {\scriptsize{$f$,$f$-$x_i$}};
\node at (0.2,-0.4) {\scriptsize{$f$-$y_1$-$y_2$,$f$-$x_i$}};
\draw  (v10) edge (u1);
\draw  (u1) edge (v6);
\draw  (v10) edge (v5);
\draw  (v5) edge (v7);
\draw  (v10) edge (v8);
\draw  (v8) edge (v4);
\node (w1) at (1.8,-0.6) {\scriptsize{$5-m$}};
\node (w2) at (1.7,-1.1) {\scriptsize{$4-m$}};
\draw (v3) .. controls (-0.9,-0.6) and (0.6,-1) .. (w2);
\draw (w2) .. controls (2.5,-1.1) and (3.3,-0.8) .. (v4);
\draw (v6) .. controls (0.2,-0.2) and (1,-0.5) .. (w1);
\draw (w1) .. controls (2.7,-0.5) and (3.3,-0.2) .. (v4);
\end{tikzpicture}
\ee
which describes the $5d$ gauge theory $\fe_7+\left(\frac92-m\right)\F$ for $2\le m\le 4$. Studying the Calabi-Yau threefold along the direction (\ref{D}), we find that it describes a $5d$ SCFT for $2\le m\le 4$.

\section{Criterion for surface decoupling}\label{decouple}
In this section, we will discuss a general procedure for decompactifying a compact surface\footnote{From now on, we will refer to 2-cycles as curves and 4-cycles as surfaces.} $S_i$ in a Calabi-Yau threefold $X$. We would like to decompactify $S_i$ such that the rest of the Calabi-Yau threefold remains compact. This means that the gluing curves $C_{ij}^\alpha$ in $S_i$ which glue $S_i$ to $S_j$ must remain compact for all $j\neq i$ and for all $\alpha$. In addition to these curves, we might also consider keeping some other curves $C_i^\alpha$ in $S_i$ compact during the decompactification process. 

Let $\cM$ be the Mori cone of $S_i$ and let $\cC_0$ be the sub-cone generated by non-negative linear combinations of the curves $C_{ij}^\alpha$ and $C_i^\alpha$. Now, consider a curve $C$ in $\cC_0$. It will in general admit multiple decompositions of the form
\be\label{d}
C=\sum n_\mu C_\mu
\ee
where $C_\mu$ are generators of the Mori cone of $S_i$ and $n_\mu\ge0$. Since the volume of all $C_\mu$ must remain non-negative throughout the process and the volume of $C$ must remain finite, it follows that the volume of any $C_\mu$ appearing in any of the above decompositions of $C$ (with $n_\mu\neq0$) must remain finite. In this way, considering all $C\in\cC_0$, we find a set $\cS_0$ of Mori cone generators which must remain compact but which are not in $\cC_0$. 

Let $\cC_1$ be the sub-cone of $\cM$ generated by the curves in $\cC_0$ and $\cS_0$. We again study decompositions of the form (\ref{d}) for every curve $C\in\cC_1$. This leads us to a set of Mori cone generators $\cS_1$ which must remain compact but are not in $\cC_1$. Joining $\cC_1$ and $\cS_1$, we obtain another sub-cone $\cC_2$ of the Mori cone. This process will converge at some step $r$ where $\cS_r$ will be empty. Then, $\cC_r\subseteq\cM$ is the sub-cone containing all the curves that must remain compact during the decompactification process.

If $\cC_r=\cM$, then it is not possible to decompactify $S_i$ while keeping the rest of the Calabi-Yau threefold compact. If $\cC_r$ is a proper subset of $\cM$, then we can decompactify the curves in $\cM-\cC_r$ thus decompactifying the surface $S_i$.

\subsection{Examples}
Let us study some examples of decoupling surfaces using the above criteria:
\ben
\item First, let us consider decoupling $S_0$ in (\ref{f41}) whose Mori cone is generated by $e$ and $f$. The only gluing curve is $h$ which can be written as
\be
h=e+f
\ee
implying that both the generators must remain compact, and thus it is impossible to decompactify $S_0$ while keeping the rest of the Calabi-Yau threefold (\ref{f41}) compact.
\item Now, let us consider the Calabi-Yau threefold associated to the untwisted compactification of the $6d$ SCFT carrying $\sp(1)$ on $-1$ curve
\be
\begin{tikzpicture} [scale=1.9]
\node (v1) at (-2.9,0) {$\mathbf{1^{10}_{1}}$};
\node at (-4.5,0.1) {\scriptsize{$2e$+$f$}};
\node at (-3.4,0.1) {\scriptsize{$2h$-$\sum x_i$}};
\node (v9) at (-4.9,0) {$\mathbf{0_{0}}$};
\draw  (v9) edge (v1);
\end{tikzpicture}
\ee
Flop $n$ blowups in $S_1$ to obtain
\be\label{sp1}
\begin{tikzpicture} [scale=1.9]
\node (v1) at (-2.8,0) {$\mathbf{1^{10-n}_{1}}$};
\node at (-4.5,0.1) {\scriptsize{$2e$+$f$-$\sum x_i$}};
\node at (-3.4,0.1) {\scriptsize{$2h$-$\sum x_i$}};
\node (v9) at (-5.1,0) {$\mathbf{0^n_{0}}$};
\draw  (v9) edge (v1);
\end{tikzpicture}
\ee
We claim that we cannot decouple $S_0$ until $n\ge4$. To see this, notice that we can write the gluing curve as
\be\label{D}
2e+f-\sum x_i=e+\left(e+f-\sum x_i\right)
\ee
Since
\be
\left(e+f-\sum x_i\right)^2\ge-1
\ee
for $n<4$, $e+f-\sum x_i$ exists in the Mori cone and hence (\ref{D}) is a valid decomposition, from which we learn that $e$ must remain compact. Now notice that
\be
e=(e-x_i)+x_i
\ee
for all blowups $x_i$, implying that $e-x_i$ and $x_i$ must remain compact. For $n=3$, we can also write the gluing curve as
\be
2e+f-\sum x_i = (e-x_1)+(e-x_2)+(f-x_3)
\ee
We can also permute $x_1,x_2,x_3$ on the right hand side of the above equation. Thus, $f-x_i$ for all $x_i$ must also remain compact. Similarly, the reader can show that all $f-x_i$ must remain compact for the $n=1,2$ cases as well, and $f$ must remain compact for the $n=0$ case. This exhausts all the generators of the Mori cone and we find that it is impossible to decouple $S_0$ for $n<4$.

\ni For $n=4$, we must keep $e+f-x_i-x_j-x_k$ compact for any three distinct blowups $x_i,x_j,x_k$. We must also keep $e-x_i$ compact for all $x_i$. But there are no restrictions on $f-x_i$ or $x_j$ and we can decompactify all of them. This leads us to the Calabi-Yau threefold
\be
\begin{tikzpicture} [scale=1.9]
\node (v1) at (-2.8,0) {$\mathbf{1^{6}_{1}}$};
\end{tikzpicture}
\ee
which describes the $5d$ gauge theory $\su(2)+6\F$. The compact curves $e+f-x_i-x_j-x_k$ and $e-x_i$ do not intersect the gluing curve $2e+f-\sum x_i$. Thus they are infinitely far separated from the $\su(2)+6\F$ theory\footnote{More concretely, the curves connecting the decoupled compact curves and the gluing curve go to infinite volume. For example, $x_i$ connects $2e_f-\sum x_i$ to $e-x_i$ since it intersects both at a single point.}. That is, they give rise to some decoupled massive states.

\ni This decompactification process can also be understood as the ungauging of a gauge algebra. To see this, notice that the Calabi-Yau threefold (\ref{sp1}) for $n=4$ admits an isomorphic description\footnote{See \cite{Bhardwaj:2019ngx,Bhardwaj:2019jtr} for more details on such isomorphisms.} as
\be
\begin{tikzpicture} [scale=1.9]
\node (v1) at (-2.9,0) {$\mathbf{1^{2+4}_{0}}$};
\node at (-4.6,0.1) {\scriptsize{$f$}};
\node at (-3.5,0.1) {\scriptsize{$f$-$\sum x_i$}};
\node (v9) at (-4.9,0) {$\mathbf{0^4_{0}}$};
\draw  (v9) edge (v1);
\end{tikzpicture}
\ee
which describes the $5d$ gauge theory with gauge algebra $\su(2)\oplus\su(2)$, a hyper in bifundamental and four hypers in fundamental of each $\su(2)$. Then the decompactification process discussed above is simply the tuning of the gauge coupling of the $\su(2)$ described by $S_0$ to zero, and the BPS string being decoupled is the BPS monopole associated to $\su(2)$. The decoupled massive BPS states can be identified with the four hypers of the $\su(2)$ which is being ungauged.

\ni Thus, according to our analysis, we cannot obtain $\su(2)+(10-n)\F$ for $0\le n\le3$ by integrating out a BPS string from the $5d$ KK theory obtained by an untwisted compactification of the $6d$ SCFT carrying $\sp(1)$ on $-1$ curve. This is consistent since it is known that $\su(2)+m\F$ is UV complete only for $m\le 8$.
\item In our final example, we will discuss a $5d$ KK theory which can flow to other $5d$ KK theories upon integrating out BPS strings. In the above two examples, the theory obtained after the flow was a $5d$ SCFT rather than a $5d$ KK theory. The $5d$ KK theory we will discuss arises via untwisted compactification of the $6d$ SCFT carrying $\so(7)$ on $-1$ curve. The associated Calabi-Yau threefold (in one of its flop frames) is
\be\label{so7}
\begin{tikzpicture} [scale=1.9]
\node (v1) at (-2.5,1.1) {$\mathbf{1^{6+2}_7}$};
\node (v2) at (-2.5,-0.5) {$\mathbf{2_{1}}$};
\node (v3) at (0.1,-0.5) {$\mathbf{3_{6}^{6}}$};
\draw  (v1) edge (v2);
\draw  (v2) edge (v3);
\node at (-1.8,0.9) {\scriptsize{$f$-$x_i$}};
\node at (-2.6,0.77) {\scriptsize{$e$}};
\node at (-2.8,-0.2) {\scriptsize{$h+2f$}};
\node at (-0.3,-0.6) {\scriptsize{$e$}};
\node at (-2.1,-0.6) {\scriptsize{$2h$}};
\node at (-0.1,-0.1) {\scriptsize{$f$-$ x_i$}};
\node (v4) at (-4.4,-0.5) {$\mathbf{0^{2}_{1}}$};
\draw  (v4) edge (v2);
\node at (-4,-0.6) {\scriptsize{$e$}};
\node at (-2.9,-0.6) {\scriptsize{$e$}};
\node (v5) at (-1.2,0.3) {\scriptsize{$6$}};
\draw  (v1) edge (v5);
\draw  (v5) edge (v3);
\node (v6) at (-3.5,0.3) {\scriptsize{$2$}};
\draw  (v4) edge (v6);
\draw  (v6) edge (v1);
\node at (-4.3,-0.1) {\scriptsize{$f$-$ x_i$}};
\node at (-3.1,0.9) {\scriptsize{$f$-$y_i$}};
\end{tikzpicture}
\ee
We would like to decompactify $S_1$, which is not possible in this flop frame. To see it, notice that the sum of the gluing curves $e$, $f-x_i$, $f-y_i$ is
\be
h+f-\sum x_i-\sum y_i=(h-\sum x_i-\sum y_i)+f
\ee
which implies that the curves $h-\sum x_i-\sum y_i$ and $f$ must remain compact. The compactness of $f$ implies that all of the $f-x_i$ and $x_j$ must remain compact. Thus all the Mori cone generators must remain compact, and the surface $S_1$ cannot be decompactified. This is good because if $S_1$ could be decompactified, we would obtain the Calabi-Yau threefold
\be
\begin{tikzpicture} [scale=1.9]
\node (v2) at (-2.5,-0.5) {$\mathbf{2_{1}}$};
\node (v3) at (-0.5,-0.5) {$\mathbf{3_{6}^{6}}$};
\draw  (v2) edge (v3);
\node at (-0.9,-0.6) {\scriptsize{$e$}};
\node at (-2.1,-0.6) {\scriptsize{$2h$}};
\node (v4) at (-4.4,-0.5) {$\mathbf{0^{2}_{1}}$};
\draw  (v4) edge (v2);
\node at (-4,-0.6) {\scriptsize{$e$}};
\node at (-2.9,-0.6) {\scriptsize{$e$}};
\end{tikzpicture}
\ee
which describes the $5d$ gauge theory $\so(7)+2\F+6\S$ \cite{Bhardwaj:2019ngx} (where $\S$ denotes a hyper in spinor representation). However, this $5d$ gauge theory exceeds the bounds presented in \cite{Jefferson:2017ahm} and hence cannot describe a $5d$ SCFT or a $5d$ KK theory.

\ni However, if we flop an $x_i$ living in $S_3$ of (\ref{so7}) to obtain
\be
\begin{tikzpicture} [scale=1.9]
\node (v1) at (-2.5,1.1) {$\mathbf{1^{5+2+2}_7}$};
\node (v2) at (-2.5,-0.5) {$\mathbf{2_{1}}$};
\node (v3) at (0.1,-0.5) {$\mathbf{3_{6}^{5}}$};
\draw  (v1) edge (v2);
\draw  (v2) edge (v3);
\node at (-1.6,0.9) {\scriptsize{$f$-$x_i,f$-$\sum z_i$}};
\node at (-2.6,0.77) {\scriptsize{$e$}};
\node at (-2.8,-0.2) {\scriptsize{$h+2f$}};
\node at (-0.3,-0.6) {\scriptsize{$e$}};
\node at (-2.1,-0.6) {\scriptsize{$2h$}};
\node at (-0.1,-0.1) {\scriptsize{$f$-$ x_i,f$}};
\node (v4) at (-4.4,-0.5) {$\mathbf{0^{2}_{1}}$};
\draw  (v4) edge (v2);
\node at (-4,-0.6) {\scriptsize{$e$}};
\node at (-2.9,-0.6) {\scriptsize{$e$}};
\node (v5) at (-1.2,0.3) {\scriptsize{$6$}};
\draw  (v1) edge (v5);
\draw  (v5) edge (v3);
\node (v6) at (-3.5,0.3) {\scriptsize{$2$}};
\draw  (v4) edge (v6);
\draw  (v6) edge (v1);
\node at (-4.3,-0.1) {\scriptsize{$f$-$ x_i$}};
\node at (-3.1,0.9) {\scriptsize{$f$-$y_i$}};
\end{tikzpicture}
\ee
then we can decompactify $S_1$ by decompactifying all $x_i,y_i,z_i$ living in $S_1$ and obtain
\be
\begin{tikzpicture} [scale=1.9]
\node (v2) at (-2.5,-0.5) {$\mathbf{2_{1}}$};
\node (v3) at (-0.5,-0.5) {$\mathbf{3_{6}^{5}}$};
\draw  (v2) edge (v3);
\node at (-0.9,-0.6) {\scriptsize{$e$}};
\node at (-2.1,-0.6) {\scriptsize{$2h$}};
\node (v4) at (-4.4,-0.5) {$\mathbf{0^{2}_{1}}$};
\draw  (v4) edge (v2);
\node at (-4,-0.6) {\scriptsize{$e$}};
\node at (-2.9,-0.6) {\scriptsize{$e$}};
\end{tikzpicture}
\ee
which describes the $5d$ gauge theory $\so(7)+2\F+5\S$ which should be a $5d$ KK theory according to \cite{Jefferson:2017ahm}. Indeed, it is shown in \cite{Bhardwaj:2020gyu} that the KK theory obtained by untwisted compactification of $6d$ SCFT carrying $\fg_2$ on $-1$ curve is described by the $5d$ gauge theory $\so(7)+2\F+5\S$.

\ni We can also flop an $x_i$ living in $S_0$ of (\ref{so7}) to obtain
\be
\begin{tikzpicture} [scale=1.9]
\node (v1) at (-2.5,1.1) {$\mathbf{1^{6+1+2}_7}$};
\node (v2) at (-2.5,-0.5) {$\mathbf{2_{1}}$};
\node (v3) at (0.1,-0.5) {$\mathbf{3_{6}^{6}}$};
\draw  (v1) edge (v2);
\draw  (v2) edge (v3);
\node at (-1.8,0.9) {\scriptsize{$f$-$x_i$}};
\node at (-2.6,0.77) {\scriptsize{$e$}};
\node at (-2.8,-0.2) {\scriptsize{$h+2f$}};
\node at (-0.3,-0.6) {\scriptsize{$e$}};
\node at (-2.1,-0.6) {\scriptsize{$2h$}};
\node at (-0.1,-0.1) {\scriptsize{$f$-$ x_i$}};
\node (v4) at (-4.4,-0.5) {$\mathbf{0^{1}_{1}}$};
\draw  (v4) edge (v2);
\node at (-4,-0.6) {\scriptsize{$e$}};
\node at (-2.9,-0.6) {\scriptsize{$e$}};
\node (v5) at (-1.2,0.3) {\scriptsize{$6$}};
\draw  (v1) edge (v5);
\draw  (v5) edge (v3);
\node (v6) at (-3.5,0.3) {\scriptsize{$2$}};
\draw  (v4) edge (v6);
\draw  (v6) edge (v1);
\node at (-4.3,-0.1) {\scriptsize{$f$-$ x,f$}};
\node at (-3.3,0.9) {\scriptsize{$f$-$y,f$-$\sum z_i$}};
\end{tikzpicture}
\ee
in which we can decompactify $S_1$ to obtain
\be
\begin{tikzpicture} [scale=1.9]
\node (v2) at (-2.5,-0.5) {$\mathbf{2_{1}}$};
\node (v3) at (-0.5,-0.5) {$\mathbf{3_{6}^{6}}$};
\draw  (v2) edge (v3);
\node at (-0.9,-0.6) {\scriptsize{$e$}};
\node at (-2.1,-0.6) {\scriptsize{$2h$}};
\node (v4) at (-4.4,-0.5) {$\mathbf{0^{1}_{1}}$};
\draw  (v4) edge (v2);
\node at (-4,-0.6) {\scriptsize{$e$}};
\node at (-2.9,-0.6) {\scriptsize{$e$}};
\end{tikzpicture}
\ee
which describes the $5d$ gauge theory $\so(7)+\F+6\S$ which is equivalent to the $5d$ KK theory obtained by twisting the $6d$ SCFT carrying $\su(4)$ on $-1$ curve by the outer automorphism of $\su(4)$ \cite{Bhardwaj:2020gyu}.
\een

\section*{Acknowledgements}
The work of the author is supported by NSF grant PHY-1719924. The author thanks Gabi Zafrir for many important discussions. The author also thanks Yuji Tachikawa and Kavli IPMU for hosting the visit during which this work was initiated.

\bibliographystyle{ytphys}
\let\bbb\bibitem\def\bibitem{\itemsep4pt\bbb}
\bibliography{ref}

\providecommand{\href}[2]{#2}\begingroup\raggedright\begin{thebibliography}{10}

\bibitem{Jefferson:2017ahm}
P.~Jefferson, H.-C. Kim, C.~Vafa, and G.~Zafrir, ``{Towards Classification of
  5d SCFTs: Single Gauge Node},''
\href{http://arxiv.org/abs/1705.05836}{{\ttfamily arXiv:1705.05836 [hep-th]}}.

\bibitem{Jefferson:2018irk}
P.~Jefferson, S.~Katz, H.-C. Kim, and C.~Vafa, ``{On Geometric Classification
  of 5d SCFTs},'' \href{http://dx.doi.org/10.1007/JHEP04(2018)103}{{\em JHEP}
  {\bfseries 04} (2018) 103},
\href{http://arxiv.org/abs/1801.04036}{{\ttfamily arXiv:1801.04036 [hep-th]}}.

\bibitem{Bhardwaj:2018yhy}
L.~Bhardwaj and P.~Jefferson, ``{Classifying 5d SCFTs via 6d SCFTs: Rank
  one},''
\href{http://arxiv.org/abs/1809.01650}{{\ttfamily arXiv:1809.01650 [hep-th]}}.

\bibitem{Bhardwaj:2018vuu}
L.~Bhardwaj and P.~Jefferson, ``{Classifying 5d SCFTs via 6d SCFTs: Arbitrary
  rank},''
\href{http://arxiv.org/abs/1811.10616}{{\ttfamily arXiv:1811.10616 [hep-th]}}.

\bibitem{Apruzzi:2018nre}
F.~Apruzzi, L.~Lin, and C.~Mayrhofer, ``{Phases of 5d SCFTs from M-/F-theory on
  Non-Flat Fibrations},'' \href{http://dx.doi.org/10.1007/JHEP05(2019)187}{{\em
  JHEP} {\bfseries 05} (2019) 187},
\href{http://arxiv.org/abs/1811.12400}{{\ttfamily arXiv:1811.12400 [hep-th]}}.

\bibitem{Apruzzi:2019vpe}
F.~Apruzzi, C.~Lawrie, L.~Lin, S.~Schafer-Nameki, and Y.-N. Wang, ``{5d
  Superconformal Field Theories and Graphs},''
\href{http://arxiv.org/abs/1906.11820}{{\ttfamily arXiv:1906.11820 [hep-th]}}.

\bibitem{Apruzzi:2019opn}
F.~Apruzzi, C.~Lawrie, L.~Lin, S.~Schafer-Nameki, and Y.-N. Wang, ``{Fibers add
  Flavor, Part I: Classification of 5d SCFTs, Flavor Symmetries and BPS
  States},''
\href{http://arxiv.org/abs/1907.05404}{{\ttfamily arXiv:1907.05404 [hep-th]}}.

\bibitem{Bhardwaj:2019fzv}
L.~Bhardwaj, P.~Jefferson, H.-C. Kim, H.-C. Tarazi, and C.~Vafa, ``{Twisted
  Circle Compactification of 6d SCFTs},''
\href{http://arxiv.org/abs/1909.11666}{{\ttfamily arXiv:1909.11666 [hep-th]}}.

\bibitem{Bhardwaj:2019jtr}
L.~Bhardwaj, ``{On the classification of $5d$ SCFTs},''
\href{http://arxiv.org/abs/1909.09635}{{\ttfamily arXiv:1909.09635 [hep-th]}}.

\bibitem{DelZotto:2017pti}
M.~Del~Zotto, J.~J. Heckman, and D.~R. Morrison, ``{6D SCFTs and Phases of 5D
  Theories},'' \href{http://dx.doi.org/10.1007/JHEP09(2017)147}{{\em JHEP}
  {\bfseries 09} (2017) 147},
\href{http://arxiv.org/abs/1703.02981}{{\ttfamily arXiv:1703.02981 [hep-th]}}.

\bibitem{Hayashi:2018bkd}
H.~Hayashi, S.-S. Kim, K.~Lee, and F.~Yagi, ``{5-brane webs for 5d $
  \mathcal{N} $ = 1 G$_{2}$ gauge theories},''
  \href{http://dx.doi.org/10.1007/JHEP03(2018)125}{{\em JHEP} {\bfseries 03}
  (2018) 125},
\href{http://arxiv.org/abs/1801.03916}{{\ttfamily arXiv:1801.03916 [hep-th]}}.

\bibitem{Hayashi:2018lyv}
H.~Hayashi, S.-S. Kim, K.~Lee, and F.~Yagi, ``{Dualities and 5-brane webs for
  5d rank 2 SCFTs},'' \href{http://dx.doi.org/10.1007/JHEP12(2018)016}{{\em
  JHEP} {\bfseries 12} (2018) 016},
\href{http://arxiv.org/abs/1806.10569}{{\ttfamily arXiv:1806.10569 [hep-th]}}.

\bibitem{Closset:2018bjz}
C.~Closset, M.~Del~Zotto, and V.~Saxena, ``{Five-dimensional SCFTs and gauge
  theory phases: an M-theory/type IIA perspective},''
  \href{http://dx.doi.org/10.21468/SciPostPhys.6.5.052}{{\em SciPost Phys.}
  {\bfseries 6} no.~5, (2019) 052},
\href{http://arxiv.org/abs/1812.10451}{{\ttfamily arXiv:1812.10451 [hep-th]}}.

\bibitem{Hayashi:2019yxj}
H.~Hayashi, S.-S. Kim, K.~Lee, and F.~Yagi, ``{Rank-3 antisymmetric matter on
  5-brane webs},'' \href{http://dx.doi.org/10.1007/JHEP05(2019)133}{{\em JHEP}
  {\bfseries 05} (2019) 133},
\href{http://arxiv.org/abs/1902.04754}{{\ttfamily arXiv:1902.04754 [hep-th]}}.

\bibitem{Fluder:2019szh}
M.~Fluder, S.~M. Hosseini, and C.~F. Uhlemann, ``{Black hole microstate
  counting in Type IIB from 5d SCFTs},''
  \href{http://dx.doi.org/10.1007/JHEP05(2019)134}{{\em JHEP} {\bfseries 05}
  (2019) 134},
\href{http://arxiv.org/abs/1902.05074}{{\ttfamily arXiv:1902.05074 [hep-th]}}.

\bibitem{Kim:2019dqn}
H.-C. Kim, S.-S. Kim, and K.~Lee, ``{Higgsing and Twisting of 6d $D_N$ gauge
  theories},''
\href{http://arxiv.org/abs/1908.04704}{{\ttfamily arXiv:1908.04704 [hep-th]}}.

\bibitem{Uhlemann:2019ypp}
C.~F. Uhlemann, ``{Exact results for 5d SCFTs of long quiver type},''
\href{http://arxiv.org/abs/1909.01369}{{\ttfamily arXiv:1909.01369 [hep-th]}}.

\bibitem{Bhardwaj:2019ngx}
L.~Bhardwaj, ``{Dualities of $5d$ gauge theories from S-duality},''
\href{http://arxiv.org/abs/1909.05250}{{\ttfamily arXiv:1909.05250 [hep-th]}}.

\bibitem{Apruzzi:2019enx}
F.~Apruzzi, C.~Lawrie, L.~Lin, S.~Schafer-Nameki, and Y.-N. Wang, ``{Fibers add
  Flavor, Part II: 5d SCFTs, Gauge Theories, and Dualities},''
\href{http://arxiv.org/abs/1909.09128}{{\ttfamily arXiv:1909.09128 [hep-th]}}.

\bibitem{Saxena:2019wuy}
V.~Saxena, ``{Rank-two 5d SCFTs from M-theory at isolated toric singularities:
  a systematic study},''
\href{http://arxiv.org/abs/1911.09574}{{\ttfamily arXiv:1911.09574 [hep-th]}}.

\bibitem{Heckman:2015bfa}
J.~J. Heckman, D.~R. Morrison, T.~Rudelius, and C.~Vafa, ``{Atomic
  Classification of 6D SCFTs},''
  \href{http://dx.doi.org/10.1002/prop.201500024}{{\em Fortsch. Phys.}
  {\bfseries 63} (2015) 468--530},
\href{http://arxiv.org/abs/1502.05405}{{\ttfamily arXiv:1502.05405 [hep-th]}}.

\bibitem{Bhardwaj:2019hhd}
L.~Bhardwaj, ``{Revisiting the classifications of $6d$ SCFTs and LSTs},''
\href{http://arxiv.org/abs/1903.10503}{{\ttfamily arXiv:1903.10503 [hep-th]}}.

\bibitem{Heckman:2013pva}
J.~J. Heckman, D.~R. Morrison, and C.~Vafa, ``{On the Classification of 6D
  SCFTs and Generalized ADE Orbifolds},''
  \href{http://dx.doi.org/10.1007/JHEP06(2015)017,
  10.1007/JHEP05(2014)028}{{\em JHEP} {\bfseries 05} (2014) 028},
  \href{http://arxiv.org/abs/1312.5746}{{\ttfamily arXiv:1312.5746 [hep-th]}}.
[Erratum: JHEP06,017(2015)].

\bibitem{Bhardwaj:2015xxa}
L.~Bhardwaj, ``{Classification of 6d $ \mathcal{N}=\left(1,0\right) $ gauge
  theories},'' \href{http://dx.doi.org/10.1007/JHEP11(2015)002}{{\em JHEP}
  {\bfseries 11} (2015) 002},
\href{http://arxiv.org/abs/1502.06594}{{\ttfamily arXiv:1502.06594 [hep-th]}}.

\bibitem{Bhardwaj:2018jgp}
L.~Bhardwaj, D.~R. Morrison, Y.~Tachikawa, and A.~Tomasiello, ``{The frozen
  phase of F-theory},'' \href{http://dx.doi.org/10.1007/JHEP08(2018)138}{{\em
  JHEP} {\bfseries 08} (2018) 138},
\href{http://arxiv.org/abs/1805.09070}{{\ttfamily arXiv:1805.09070 [hep-th]}}.

\bibitem{Seiberg:1996qx}
N.~Seiberg, ``{Nontrivial fixed points of the renormalization group in
  six-dimensions},''
  \href{http://dx.doi.org/10.1016/S0370-2693(96)01424-4}{{\em Phys.Lett.}
  {\bfseries B390} (1997) 169--171},
\href{http://arxiv.org/abs/hep-th/9609161}{{\ttfamily arXiv:hep-th/9609161
  [hep-th]}}.

\bibitem{Danielsson:1997kt}
U.~H. Danielsson, G.~Ferretti, J.~Kalkkinen, and P.~Stjernberg, ``{Notes on
  supersymmetric gauge theories in five-dimensions and six-dimensions},''
  \href{http://dx.doi.org/10.1016/S0370-2693(97)00645-X}{{\em Phys.Lett.}
  {\bfseries B405} (1997) 265--270},
\href{http://arxiv.org/abs/hep-th/9703098}{{\ttfamily arXiv:hep-th/9703098
  [hep-th]}}.

\bibitem{Bhardwaj:2020gyu}
L.~Bhardwaj and G.~Zafrir, ``{Classification of 5d $ \mathcal{N} $ = 1 gauge
  theories},'' \href{http://dx.doi.org/10.1007/JHEP12(2020)099}{{\em JHEP}
  {\bfseries 12} (2020) 099}, \href{http://arxiv.org/abs/2003.04333}{{\ttfamily
  arXiv:2003.04333 [hep-th]}}.

\end{thebibliography}\endgroup

\end{document}